\documentclass{iaus}
\usepackage{graphicx}

\title[Binarity and the morpho-kinematics of PNe] 
{The influence of binarity on the morpho-kinematics of planetary nebulae}

\author[David Jones et al.]   
{David Jones$^1$,
  Amy A. Tyndall$^2$,
  Myfanwy Lloyd$^2$
 \and Miguel Santander-Garc\'ia$^3$
 }

\affiliation{$^1$European Southern Observatory, Alonso de C\'ordova 3107, Casilla 19001, Santiago, Chile \\ email: {\tt djones@eso.org} \\[\affilskip]
$^2$Jodrell Bank Centre for Astrophysics, University of Manchester, Oxford Road, Manchester, M13 9PL, UK \\email: {\tt amy.tyndall@postgrad.manchester.ac.uk}, {\tt myfanwy.lloyd@manchester.ac.uk}\\[\affilskip]
$^3$Observatorio Astron\'omico Nacional, Ap 112, 28803 Alcal\'a de Henares, Spain \\email: {\tt m.santander@oan.es}
}

\pubyear{2010}
\volume{283}  
\pagerange{x--y}
\jname{Planetary Nebulae: an Eye to the Future}
\editors{A. Manchado, L. Stanghellini \& D. Schoenberner, eds.}
\begin{document}

\maketitle

\begin{abstract}
The role of central star binarity in the shaping of planetary nebulae (PNe) has been the subject of much debate, with single stars believed to be incapable of producing the most highly collimated morphologies. However, observational support for binary-induced shaping has been sadly lacking. Here, we highlight the results of a continuing programme to spatio-kinematically model the morphologies of all PNe known to contain a close binary central star. Spatio-kinematical modelling is imperative for these objects, as it circumvents the degeneracy between morphology and orientation which can adversely affect determinations of morphology based on imaging alone. Furthermore, spatio-kinematical modelling accurately determines the orientation of the nebular shell, allowing the theoretically predicted perpendicular alignment, between nebular symmetry axis and binary orbital plane, to be tested. To date, every PN subjected to this investigation has displayed the predicted alignment, indicating that binarity has played an important role in the formation and evolution of these nebulae. The further results from this programme will be key, not only in determining whether binary interaction is responsible for shaping the studied PNe, but also in assessing the importance of binarity in the formation and evolution of all PNe in general.
\end{abstract}

Central star binarity has long been believed to be associated with the formation of aspherical PNe \cite[(de Marco 2009)]{demarco09}, with an important prediction of all shaping theories being the perpendicular alignment between nebular symmetry axis and binary orbital plane \cite[(Mitchell 2007)]{mitchell07}.  Therefore, the comparison of nebular and binary orientation provides an obvious test of this `binary hypothesis'.  Detailed spatio-kinematic modelling is required to accurately determine the morphology and inclination of a given PN due to projection effects in imaging studies.  So far, only six PNe with well-studied binary central stars (i.e.\ with a known binary inclination) have been the subject of detailed spatio-kinematic study - but all have been found to show the expected alignment.  The details of the six PNe are listed in table \ref{tab:neb}.

A collective analysis of the morphologies and kinematics of the six PNe reveals far more differences than similarities \cite[(Jones \etal\ 2011)]{jones11}, indicating that understanding the exact shaping mechanism(s) involved will be far from trivial.  However, some intriguing avenues for study are thrown up by the sample, all of which are discussed in more detail in \cite[Jones \etal\ 2011]{jones11} and will be further explored in upcoming papers (Jones \etal{}, Santander-Garc\'ia \etal\ in prep.).  Possibly the most interesting discovery is the presence of extended outflows in three PNe which are older than the central nebula, most likely being formed before the central binary entered the common-envelope phase (see table \ref{tab:ages}).

A progamme is now underway to derive spatio-kinematic models from all PNe known to contain a post-common-envelope central star, not only to compare the nebular inclinations to those of the central stars but also to further investigate the links between the morpho-kinematic properties of the nebula and the parameters of the central star systems.  The results of this investigation will prove invaluable in understanding the processes at work in formation and evolution of these systems.

\begin{table}
\begin{center}
\caption{A table of PNe, with well studied binary central stars, that have been the subject of detailed spatio-kinematic study.}
\label{tab:neb}
{\scriptsize
\begin{tabular}{|c|c|c|}
\hline
{\bf PN} & {\bf Morphology} & {\bf References}\\ \hline
Abell 41 & Open-ended bipolar with equatorial torus & \cite[Jones \etal\ (2010)]{jones10}\\
Abell 63 & Barrel-like with bipolar ``end-caps'' & \cite[Mitchell \etal\ (2007)]{mitchell07}\\
Abell 65 & Double-shelled bipolar & \cite[Huckvale \etal\ (2011)]{huckvale11}\\
HaTr~4 & Elliptical with equatorial torus & \cite[Tyndall \etal\ (2011a,b)]{tyndall11a,tyndall11b}\\
NGC~6337 & Torus without bipolar lobes, &\cite[Garc\'ia-D\'iaz \etal\ (2009)]{garciadiaz09},\\
& jet-like corkscrew outflow &  \cite[Hillwig \etal\ (2010)]{hillwig10}\\
Sp~1 & End-on bipolar & \cite[Jones \etal\ (2011)]{jones11} \\
\hline
\end{tabular}
}
\end{center}
\end{table}

\begin{table}
\begin{center}
\caption{The kinematical ages of PNe with binary central stars and their polar outflows.}
\label{tab:ages}
{\scriptsize
\begin{tabular}{|c|c|c|c|}
\hline
{\bf PN} & {\bf PN age} & {\bf Polar outflow age} & {\bf Reference}\\
& {\bf (yrs kpc$^{-1}$} & {\bf (yrs kpc$^{-1}$} &\\ \hline
Abell 63 & 3500$\pm$200 & 5200$\pm$1200 & \cite[Mitchell \etal\ (2007)]{mitchell07}\\
ETHOS 1 & 900$\pm$100 & 1750$\pm$250 & \cite[Miszalski \etal\ (2011)]{miszalski11}\\
The Necklace & 1100$\pm$100 & 2350$\pm$450 & \cite[Corradi \etal\ (2011)]{corradi11}\\ 
\hline
\end{tabular}
}
\end{center}
\end{table}

\end{document}